\documentclass[aps,prl,twocolumn,superscriptaddress,showpacs,letter]{revtex4}

\usepackage{graphicx}
\usepackage{dcolumn}
\usepackage{bm}
\usepackage{cancel}
\usepackage{color}

\newcommand{\undoped}{BaFe$_2$As$_2$}
\newcommand{\Co}{Ba(Co$_{0.07}$Fe$_{0.93}$)$_2$As$_2$}
\newcommand{\AFS}{A$_{x}$Fe$_{2-y}$Se$_2$}
\newcommand{\KFS}{K$_{x}$Fe$_{2-y}$Se$_2$}
\newcommand{\RFS}{Rb$_{x}$Fe$_{2-y}$Se$_2$}
\newcommand{\EF}{$E_F$}
\newcommand{\kf}{$k_F$}
\newcommand{\dxy}{$d_{xy}$}
\newcommand{\dxz}{$d_{xz}$}
\newcommand{\dyz}{$d_{yz}$}
\newcommand{\dxxyy}{$d_{x^2-y^2}$}
\newcommand{\dz}{$d_{3z^2-r^2}$}
\newcommand{\dxzyz}{$d_{xz}$/$d_{yz}$}

\newcommand{\mub}{$\mu_{\tiny{\textrm{B}}}$}

\begin{document}

\title{Observation of Temperature-Induced Crossover to an Orbital-Selective Mott Phase in \AFS~(A=K, Rb) Superconductors}

\author{M. Yi}
\affiliation{Stanford Institute of Materials and Energy Sciences, Stanford University, Stanford, CA 94305, USA}
\affiliation{Departments of Physics and Applied Physics, and Geballe Laboratory for Advanced Materials, Stanford University, Stanford, CA 94305, USA}
\author{D.H. Lu}
\affiliation{Stanford Synchrotron Radiation Lightsource, SLAC National Accelerator Laboratory, Menlo Park, CA 94025, USA}
\author{R. Yu}
\affiliation{Department of Physics and Astronomy, Rice University, Houston, TX 77005, USA}
\author{S. C. Riggs}
\author{J.-H. Chu}
\affiliation{Stanford Institute of Materials and Energy Sciences, Stanford University, Stanford, CA 94305, USA}
\affiliation{Departments of Physics and Applied Physics, and Geballe Laboratory for Advanced Materials, Stanford University, Stanford, CA 94305, USA}
\author{B. Lv}
\affiliation{Department of Physics, Texas Center for Superconductivity, University of Houston, Houston TX 77204, USA}
\author{Z. Liu}
\affiliation{Stanford Institute of Materials and Energy Sciences, Stanford University, Stanford, CA 94305, USA}
\affiliation{Departments of Physics and Applied Physics, and Geballe Laboratory for Advanced Materials, Stanford University, Stanford, CA 94305, USA}
\author{M. Lu}
\affiliation{Stanford Institute of Materials and Energy Sciences, Stanford University, Stanford, CA 94305, USA}
\affiliation{National Laboratory of Solid-State Microstructures and Department of Materials Science and Engineering, Nanjing University, Nanjing 210093, P. R. China}
\author{Y.-T. Cui}
\affiliation{Stanford Institute of Materials and Energy Sciences, Stanford University, Stanford, CA 94305, USA}
\author{M. Hashimoto}
\affiliation{Stanford Synchrotron Radiation Lightsource, SLAC National Accelerator Laboratory, Menlo Park, CA 94025, USA}
\author{S.-K. Mo}
\author{Z. Hussain}
\affiliation{Advanced Light Source, Lawrence Berkeley National Lab, Berkeley, CA 94720, USA}
\author{C. W. Chu}
\affiliation{Department of Physics, Texas Center for Superconductivity, University of Houston, Houston TX 77204, USA}
\author{I.R. Fisher}
\affiliation{Stanford Institute of Materials and Energy Sciences, Stanford University, Stanford, CA 94305, USA}
\affiliation{Departments of Physics and Applied Physics, and Geballe Laboratory for Advanced Materials, Stanford University, Stanford, CA 94305, USA}
\author{Q. Si}
\affiliation{Department of Physics and Astronomy, Rice University, Houston, TX 77005, USA}
\author{Z.-X. Shen}
\affiliation{Stanford Institute of Materials and Energy Sciences, Stanford University, Stanford, CA 94305, USA}
\affiliation{Departments of Physics and Applied Physics, and Geballe Laboratory for Advanced Materials, Stanford University, Stanford, CA 94305, USA}

\date{\today}

\begin{abstract}
In this work, we study the \AFS~(A=K, Rb) superconductors using angle-resolved photoemission spectroscopy. In the low temperature state, we observe an orbital-dependent renormalization for the bands near the Fermi level in which the \dxy~bands are heavily renormliazed compared to the \dxzyz~bands. Upon increasing temperature to above 150K, the system evolves into a state in which the \dxy~bands have diminished spectral weight while the \dxzyz~bands remain metallic. Combined with theoretical calculations, our observations can be consistently understood as a temperature induced crossover from a metallic state at low temperature to an orbital-selective Mott phase (OSMP) at high temperatures. Furthermore, the fact that the superconducting state of \AFS~is near the boundary of such an OSMP constraints the system to have sufficiently strong on-site Coulomb interactions and Hund's coupling, and hence highlight the non-trivial role of electron correlation in this family of iron superconductors.
\end{abstract}

\pacs{74.25.Jb, 74.70.-b, 79.60.-i}
\maketitle

Electron correlation remains a central focus in the study of high temperature superconductors. The strongly correlated cuprate superconductors are understood as doped Mott insulators (MI)~\cite{1LeePA06} while the iron-based superconductors (FeSCs) have been found to be at most moderately correlated~\cite{2LuDH08,3YangWL09,4Qazibash09}. Even though the low energy electronic structures of different FeSCs families share the common Fe $3d$ bands, there are systematic variations in their physical properties, such as ordered magnetic moment and effective mass~\cite{5YinZP11}. In general, electron correlation is the weakest in iron phosphides with relatively low mass renormalization~\cite{2LuDH08}, and moderate in the more extensively studied iron arsenides~\cite{2LuDH08,3YangWL09}. The Fe(Te,Se) chalcogenide family, in comparison, seems to harbor stronger correlation as inferred from larger ordered moment, yet metallic resistivity is still observed~\cite{6Liu10}. The newest iron chalcogenide superconductors, \AFS~(A=alkali metal)~\cite{7Guo10,8Krzton11,9Li11,10Fang11,11Wang11} (AFS) hints at even stronger correlation with a large observed moment of 3.3\mub~\cite{12Wei11} and insulating transport behavior in the phase diagram.

Another important factor in understanding the FeSCs lies in their multi-orbital nature. In such a system, orbital-dependent behavior as well as competition between inter- and intra-orbital interactions could play a critical role in determining their physical properties. One example is the orbital anisotropy that onsets with the in-plane symmetry breaking structural transition as observed in the underdoped arsenides~\cite{13Yi11}. Another example is the different pairing symmetry that could arise considering the relative strength of inter- and intra-orbital interactions extensively studied theoretically~\cite{14chubukov12}. Theoretical models have considered correlation effects in the bad metal regime in terms of an incipient Mott picture~\cite{5YinZP11,15si08}, and the proximity to the Mott transition may be orbital-dependent even for orbitally-independent Coulomb interactions~\cite{16Yu11,17Yu11b,18Zhou11,19craco11}. What arises from the model is an orbital selective Mott phase (OSMP), in which some orbitals are Mott-localized while others remain itinerant. First introduced in the context of the Ca$_{2-x}$Sr$_x$RuO$_4$ system, an OSMP may result from both the orbital-dependent kinetic energy and the combined effects of the Hund's coupling and crystal level splittings~\cite{20anisimov02,21demedici}. An OSMP links naturally with models of coexisting itinerant and localized electrons that have been proposed to compensate for the shortcomings of both strong coupling and weak coupling approaches~\cite{22You11,23moon10}. However, to date, there has been no experimental evidence for OSMP in any FeSC.

\begin{figure*}[t]
\includegraphics[width=0.9\textwidth]{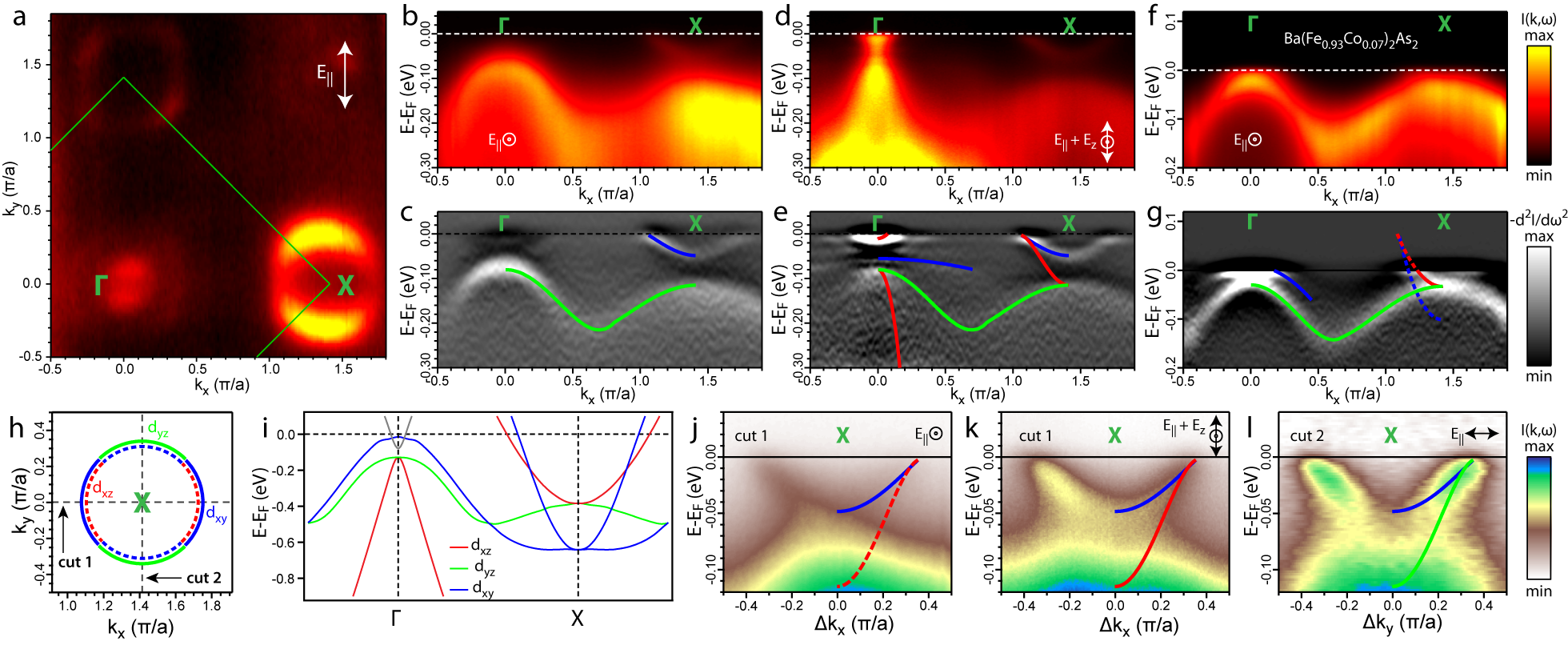}
\caption{\label{fig:fig1}Measured electronic structure of \KFS. (a) Fermi surface mapping by integrating 20meV window about \EF. Green lines outline the 2-Fe Brillouin zone. (b),(d) Spectral images and (c),(e) second derivatives taken along the $\Gamma$-X direction using light polarizations as marked. (f)-(g), equivalent of that of (b)-(c) for \Co. (h) Schematic of the dominant orbital characters of the two electron pockets of the same size near X point, with one of the pockets (dotted) imploded for clarity. (i) LDA calculations~\cite{27nekrasov11} for KFS with the dominant orbital characters labeled. (j)-(l), Spectral images taken across the X-point under different polarizations and cut directions. Guides to eye for the observable bands are overlaid, with colors indicating the dominant orbital characters-blue: \dxy; red: \dxz; green: \dyz. All data taken at 30K, with 47.5eV photons except (d), (e), and (k), which were taken with 26eV photons.}
\end{figure*}

In this paper, we present angle-resolved photoemission spectroscopy (ARPES) data from two superconducting AFSs, \KFS~(KFS) and \RFS~(RFS), with $T_C$ of 32K and 31K, respectively, as well as insulating and intermediate dopings (see SI). We observe the superconducting AFSs undergoing a temperature-induced crossover from a metallic state in which all three $t_{2g}$ orbitals-\dxy, \dxz~and \dyz~are present near the Fermi level (\EF) to a state in which the \dxy~bands has diminished spectral weight while the \dxzyz~bands remain metallic. In addition, the intermediate doping shows stronger correlation than the superconducting doping, as seen in the further renormalization of the \dxzyz~bands and the much more suppressed \dxy~intensity, while the insulating doping has no spectral weight near \EF~in any orbitals. From comparison with our theoretical calculations, the ensemble of our observations are most consistent with the understanding that the presence of strong Coulomb interactions and Hund's coupling places the superconducting AFSs near an OSMP at low temperatures, and crosses over into the OSMP via raised temperature, while the intermediate and insulating compounds are on the boundary of the OSMP and in the MI phase, respectively, suggesting the importance of electron correlation in this family of FeSCs.

\begin{figure*}
\includegraphics[width=0.95\textwidth]{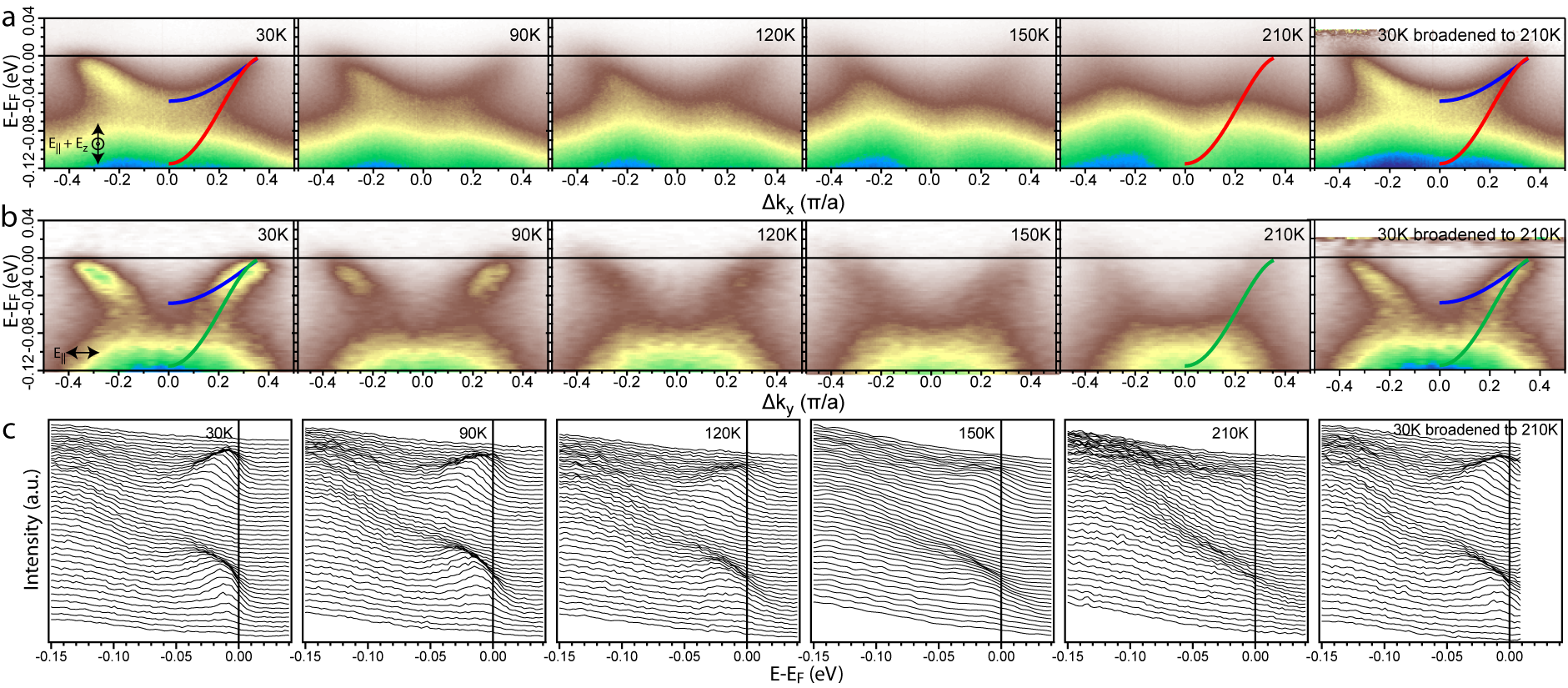}
\caption{\label{fig:fig2}Temperature dependence of electron bands near X. (a) Spectral images taken with 26eV photons along cut1. (b) Spectral images and (c) corresponding energy distribution curves (EDCs) taken with 47.5eV photons along cut2. Polarizations are as marked. Spectral images for selected temperatures are shown for both geometries. To rule out a trivial thermal broadening for the diminishing spectral weight, we artificially introduce a 210K thermal broadening to the 30K data and plot them next to the 210K data for comparison.}
\end{figure*}

The low temperature electronic structure of \KFS is shown in Fig.~\ref{fig:fig1}. The Fermi surface (FS) of KFS consists of large electron pockets at the Brillouin zone (BZ) corner-X point, and a small electron pocket at the BZ center-$\Gamma$ point (Fig.~\ref{fig:fig1}(a)), consistent with previous ARPES reports~\cite{24zhang11,25qian11,26mou11}. For the crystallographic 2-Fe unit cell, LDA calculations predict two electron pockets at the X point~\cite{27nekrasov11} (Fig.~\ref{fig:fig1}(i)). While FS map appears to show only one electron pocket at X, measurements under different polarizations (Fig.~\ref{fig:fig1}(j)-(l)) reveal the expected two electron bands with nearly degenerate Fermi crossings (\kf) but different band bottom positions-a shallower one around -0.05eV and a deeper one that extends to the top of the \dxzyz~hole-like bands at -0.12eV. The Luttinger volume of the two electron pockets gives $\sim$0.16 electrons per Fe. Considering that the C$_4$ symmetry of the crystal dictates degeneracy of the \dxzyz~electron band bottom and hole band top at X, the shallower electron band that is not degenerate with the hole-like band at higher binding energy is most likely of \dxy~character, and the deeper one \dxz~along $\Gamma$-X and \dyz~along the perpendicular direction. This observed orbital character assignment seems to contradict the LDA prediction of the shallower electron band as \dxzyz~and the deeper one \dxy~in FeSC~\cite{28graser10} (Fig.~\ref{fig:fig1}(i)), as observed in Co-doped \undoped~(Fig.~\ref{fig:fig1}(f)-(g)). However, this assignment can be understood if we consider the KFS band structure as a whole. Three filled hole bands are observed near the $\Gamma$ point (Figs.~\ref{fig:fig1}(d)-(e)), where the two lower ones can be identified as \dxzyz~and the higher one \dxy. Interestingly, the \dxy~band is far more renormalized ($\sim$a factor of 10) compared to LDA than the \dxzyz~bands ($\sim$a factor of 3), indicating stronger correlation for the \dxy~orbital. This is in contrast to the Co-doped \undoped~band structure, in which all orbitals are renormalized by roughly the same factor ($\sim$2) compared to LDA. Hence, our assignment of \dxy~character to the shallower electron band is consistent with strong orbital-dependent renormalization, which brings the deeper \dxy~electron band at X predicted by LDA to be shallower than the \dxzyz~band. This orbital-dependent renormalization behavior also emerges from our theoretical calculations as will be discussed later.

As the electronic structure at \EF~is dominated by the large electron pockets at X, we focus on these features. Fig.~\ref{fig:fig2} shows a temperature-dependent study of these bands taken in two polarization geometries. In Fig.~\ref{fig:fig2}(a), the matrix element for \dxy~is stronger than \dxz. At low temperatures, \dxy~band is clearly resolved with weaker intensity for the \dxz~band. With increasing temperature, the spectral weight of the \dxy~band noticeably diminishes, eventually revealing the remaining \dxz~band at high temperatures. In Fig.~\ref{fig:fig2}(b), the deeper \dyz~band has more intensity than the corresponding \dxz~band in Fig.~\ref{fig:fig2}(a) while the presence of the \dxy~band is still very noticeable from the increased intensity where they significantly overlap, as well as from the energy distribution curves (EDCs) shown in Fig.~\ref{fig:fig2}(c). With increasing temperature, again, the spectral weight of the \dxy~band diminishes, leaving only the \dyz~band at high temperatures, which clearly has a deeper band bottom than the shallow \dxy~band (Fig.~\ref{fig:fig2}(c)). In addition, we have artificially introduced a 210K thermal broadening to the 30K spectra as shown in the last column of Fig.~\ref{fig:fig2}. The clear contrast to the 210K data rules out a trivial thermal broadening as an origin for the observed diminishing of \dxy~spectral weight.

\begin{figure*}
\includegraphics[width=0.95\textwidth]{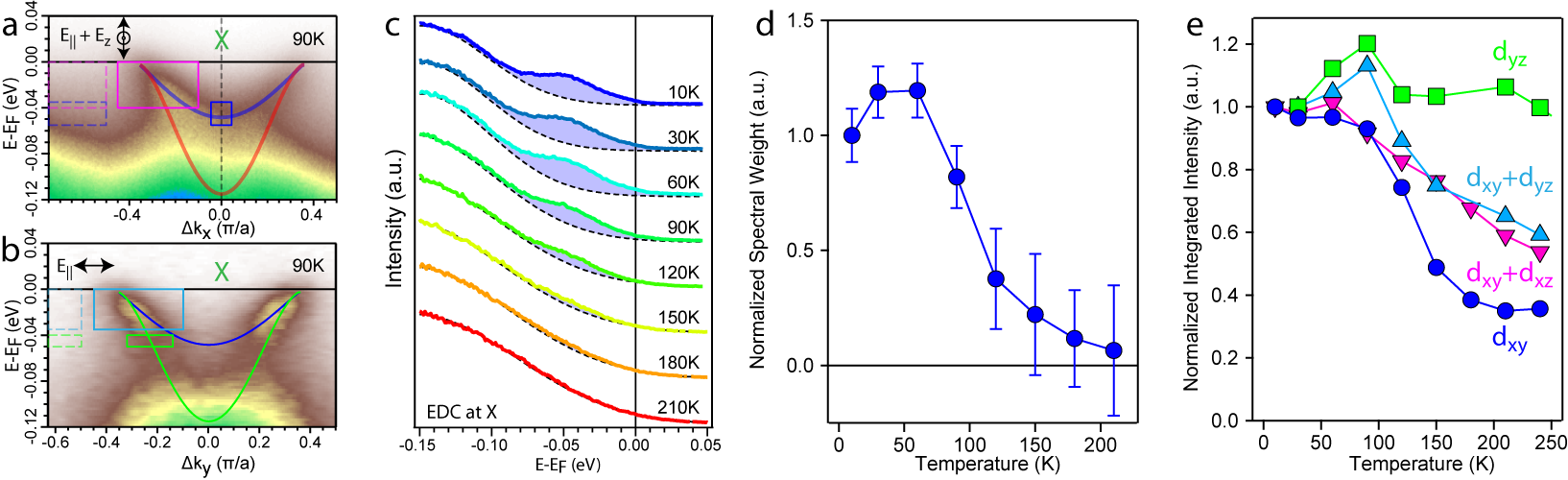}
\caption{\label{fig:fig3}Quantitative analysis of the temperature dependence. (a),(b) 90K spectral image from Fig.~\ref{fig:fig2}(a),(b). (c) EDCs at the X point from (a) for selected temperatures. Curves are offset for clarity. Each curve is fitted to a Gaussian (dotted line) for the background feature at high binding energy and a Lorentzian (shaded region) for the feature around -0.05eV. (d) The integrated Lorentzian spectral weight are plotted as a function of temperature. (e) Temperature-dependence of the averaged intensity in the regions marked by colored boxes in (a)-(b), with background (dotted box of the same energy window) for each box subtracted. The resulting temperature-dependent curve is then normalized by the initial value. The blue (green) region has dominant spectral weight from the \dxy~(\dyz) band, whereas the magenta (cyan) region is of mixed \dxy~and \dxz~(dyz) characters.}
\end{figure*}

To capture this behavior quantitatively, we analyze the temperature dependence of the EDCs at the X-point, stacked in Fig.~\ref{fig:fig3}(c). At all temperatures, there is a large hump background corresponding to the large hole-like dispersion at high binding energy. At low temperatures, there is an additional peak around -0.05eV corresponding to the \dxy~band bottom. We fit these EDCs with a Gaussian background for the large hump feature and a Lorentzian peak for the \dxy~band. The integrated spectral weight for the fitted \dxy~peak is plotted in Fig.~\ref{fig:fig3}(d), which decreases toward zero with increasing temperature, seen as a non-trivial drop between 100K and 200K. As an independent check against trivial thermal effect, we choose small regions in the raw spectral image (marked in Fig.~\ref{fig:fig3}(a)-(b)) dominated by \dxy~(blue), \dyz~(green), and mixed \dxy~and \dxzyz~(magenta/cyan) characters and plot their integrated intensities as a  function of temperature (Fig.~\ref{fig:fig3}(e)). The spectral weight from \dxy-dominated region rapidly decreases, consistent with the fitted result in Fig.~\ref{fig:fig3}(d), while that of \dyz-dominated region does not drop in a similar manner. The regions of mixed orbitals show a slower diminishing spectral weight compared to that for \dxy, reflecting the combined contributions from both \dxzyz~and \dxy~orbitals. Although this method has the uncertainty of small leakage of other orbitals into the chosen regions, which is the cause of the finite residual value for the \dxy~curve, the contrasting behavior of the \dxy~versus \dxzyz~orbitals is clearly demonstrated. A temperature cycle test was performed to exclude the possibility of sample aging (see SI). Measurements on the sister compound RFS reveal similar behavior (SI), suggesting the generality in this family of superconductors. This observation of a selected orbital that loses coherent spectral weight while the others remain metallic is reminiscent of a crossover into an OSMP in which selected orbitals become Mott localized while others remain metallic.

\begin{figure*}
\includegraphics[width=0.95\textwidth]{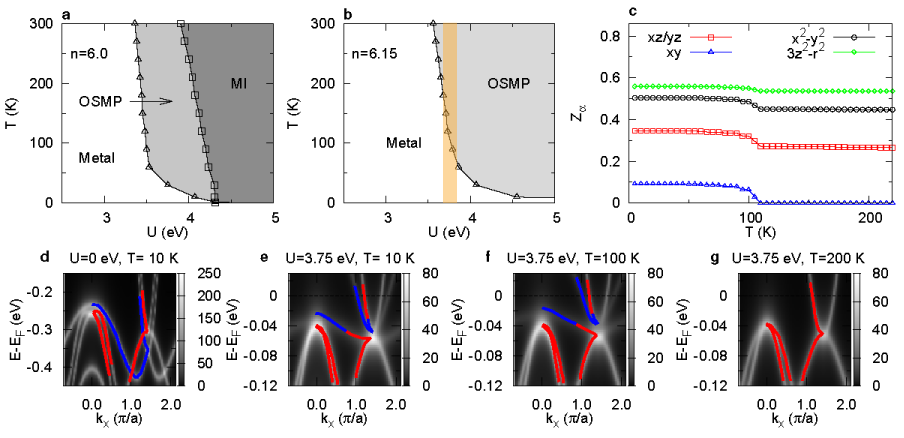}
\caption{\label{fig:fig4}Theoretical calculation based on a five-orbital Hubbard model. (a),(b) Slave-spin mean-field phase diagrams of the five-orbital Hubbard model for KFS at J/U=0.15 and two electron densities. OSMP and MI refer to orbital-selective Mott phase and Mott insulator, respectively. (c) The evolution of the orbital-resolved quasiparticle spectral weight Z$_{\alpha}$ with temperature at n=6.15, U=3.75eV, and J/U=0.15. (d)-(g) show the quasiparticle spectral functions along the $\Gamma$-X direction at n=6.15. The color curves highlight the dominant orbital of the band, with \dxzyz~in red and \dxy~in blue.}
\end{figure*}

To further understand this phenomenon, we perform theoretical calculations based on a five-orbital Hubbard model to study the metal-to-insulator transition (MIT) in the paramagnetic phase using a slave-spin mean-field method~\cite{30yu,31demedici05}. At commensurate electron filling n=6 per Fe (corresponding to Fe2$^+$ of the parent FeSC), we find the ground state of the system to be a metal, an OSMP or a MI depending on the intra-orbital repulsion U and the Hund's coupling J. Furthermore, the MIT can be triggered by increasing temperature (Fig.~\ref{fig:fig4}(a)) due to the larger entropy of the insulating phase. At a fixed interaction strength (within a certain range, Fig.~\ref{fig:fig4}(a)), the system goes from a metal to an OSMP and then to a MI with increasing temperature. The MI phase is suppressed by electron doping. By contrast, the OSMP can survive, as shown in Fig.~\ref{fig:fig4}(b) for n=6.15, which roughly corresponds to the filling of the superconducting state from ARPES measurements. From the evolution of the orbitally resolved quasiparticle spectral weight, $Z_{\alpha}$, as a function of the 
temperature (Fig.~\ref{fig:fig4}(c)), we show that the OSMP corresponds to the \dxy~orbital being Mott localized ($Z=0$) and the rest of the $3d$ orbitals remaining delocalized ($Z>0$). This result originates primarily from a combined effect of the orbital dependence of the projected density of states and the interplay between the Hund's coupling and crystal level splitting (see SI and Ref.~\onlinecite{29yu}).


To compare with the ARPES measurements, we have calculated the quasiparticle spectral function A(k,E) in the 2-Fe BZ. At low temperature and in the non-interacting limit U=0 (Fig.~\ref{fig:fig4}(d)), the electronic structure of the model agrees well with that from LDA, with the \dxy~band deeper than the \dxzyz~band at X. This order switches with sufficiently strong interaction. At U=3.75 eV (Fig.~\ref{fig:fig4}(e)), the \dxy-dominated bands are pushed above their \dxzyz~counterparts by the strongly orbital-dependent mass renormalization, as reflected in the orbital dependent quasiparticle spectral weights (Fig.~\ref{fig:fig4}(c)). The mass renormalization is the largest for the \dxy~orbital ($\sim$10), and smaller for the \dxzyz~orbitals ($\sim$3), which is compatible with the low-temperature ARPES results (Fig.~\ref{fig:fig1}(e)). The temperature-induced crossover to the OSMP is clearly seen from the suppression of the spectral weights in the \dxy~orbital that accompanies the reduction of the weights in the other orbitals (Fig.~\ref{fig:fig4}(c),(e)-(g)), in agreement with the ARPES results (Fig.~\ref{fig:fig3}).

The temperature-induced nature of the crossover constrains these AFS superconductors to be very close to the boundary of the OSMP in the zero temperature ground state, which is also the superconducting state. The best agreement between theory and experiments is achieved at U$\sim$3.75 eV. While the absolute value of this interaction is sensitive to the parameterization of the crystal levels and Hund's coupling, it is instructive to make a qualitative comparison with the case of the iron pnictides. The enhanced correlation effects for AFS tracks the reduction of the width of the (U=0) \dxy~band, which is about 0.7 of its counterpart in 1111 iron arsenides.

One known concern for the AFS materials is the existence of mesoscopic phase separation into superconducting and insulating regions~\cite{32li12}, which would both contribute spectral intensity in ARPES data. From measurement on an insulating RFS sample (Fig.~\ref{fig:figsi4}(d)), we see negligible spectral weight and no well-defined dispersions within 0.1eV from \EF, as expected for an insulator. Hence, the insulating regions in the superconducting compounds do not contribute spectral weight to the near-\EF~energy range, in which the temperature-induced crossover is observed. We have also measured a KFS sample whose resistivity is intermediate between superconducting and insulating (Fig.~\ref{fig:figsi4}(b)), and was previously proposed to be semiconducting containing both metallic and insulating phases~\cite{33chen11}. Interestingly, its \dxzyz~bands, which must come from the metallic phase, appear further renormalized by a factor of 1.3 compared with those of the superconducting compounds. In addition, we resolve small but finite spectral weight for a very shallow \dxy~electron band near the X-point (Fig.~\ref{fig:figsi5}(c)). As expected, the peak position is even closer to \EF, consistent with additional renormalization for the shallow band near X, which is harder to discern with temperature dependent study. These observations are consistent with the OSMP picture in that that the metallic phase in this KFS compound is likely even closer to the boundary of OSMP at low temperatures from the mass-diverging behavior of the \dxy~bands. For the same interaction strength, calculation from our model also identifies the low temperature ground state of the superconducting, intermediate, and insulating phases to be located close to an OSMP, just at the boundary of an OSMP, and in a MI phase, respectively (see SI).

While we cannot completely rule out alternative explanations for the observations presented above, the consistency of the totality of the observations-including strongly orbital dependent band renormalization for \dxy~versus \dxzyz~at both $\Gamma$ and X points in the low temperature metallic state, the non-trivial temperature-dependent spectral weight change for only the \dxy~band, systematic doping dependence of the related effects in the intermediate and insulating compounds-and the theoretical calculations makes this understanding a most likely scenario, suggesting that the superconductivity in this AFS family exists in close proximity to Mott behavior.

\bigskip
\begin{acknowledgments}
We thank V. Brouet, W. Ku, B. Moritz and I. Mazin for helpful discussions. ARPES experiments were performed at the Stanford Synchrotron Radiation Lightsource and the Advanced Light Source, which are both operated by the Office of Basic Energy Science, U.S. Department of Energy. The work at Stanford is supported by DOE Office of Basic Energy Science, Division of Materials Science and Engineering, under contract DE-AC02-76SF00515. The work at Rice has been supported by NSF Grant DMR-1006985 and the Robert A. Welch Foundation Grant No. C-1411. The work at Houston is supported in part by US Air Force Office of Scientific Research contract FA9550-09-1-0656, and the state of Texas through the Texas Center for Superconductivity at the University of Houston. MY thanks the NSF Graduate Research Fellowship Program for financial support.
\end{acknowledgments}
\bigskip


\setcounter{figure}{0}
\makeatletter
\renewcommand{\thefigure}{S\@arabic\c@figure}

\section{Supplementary Information}
\subsection{Materials and methods}
Single crystals of KFS were grown by the following method. Precursor FeSe was synthesized using Se powder (Alfa 99.999$\%$) and Fe powder (99.998$\%$) in a 1:1 molar ratio.  The reagents were weighed and placed in a 2mL alumina crucible.  The quartz tube was sealed after being flushed with argon and evacuated. The sealed quartz tube was placed in a furnace and heated from room temperature to 1050$^\circ$C in 12 hours. The furnace remained at 1050$^\circ$C for 12 hours, then was shut off and cooled to room temperature. Single crystals of KFS were obtained by a self-flux method with mixtures of K (Alfa, 99.95$\%$), and FeSe in molar ratios of 1:3 respectively. Potassium has a low boiling point so a small amount of additional K needs to be incorporated into the growth during synthesis.  The reagents were weighed and placed in a 2mL alumina crucible, which was then sealed in a 2mL quartz tube after being flushed with argon and evacuated. The 2mL quartz tube was then placed into a larger 5ml quartz tube and sealed after again being flushed with argon and evacuated.  The double sealed quartz tube technique is employed because potassium attacks quartz.  The sealed quartz tubes were placed in a furnace and heated from room temperature to 1040$^\circ$C over the course of 4 hours. After dwelling for 2 hours, the furnace was cooled to 850$^\circ$C.  The quartz tube was then removed from the furnace, rotated 180$^\circ$C, and spun in a centrifuge for a few seconds to separate the self-flux from the single crystals. Crystals with dimensions up to approximately 3mm x 3mm x 50mm could readily be extracted from the crucible. The crystals have a platelike morphology, with the c-axis perpendicular to the plane. Two types of KFS were studied, one superconducting with TC onset of 32K and chemical composition as K$_{0.76}$Fe$_{1.72}$Se$_2$, and the other non-superconducting with composition as K$_{0.76}$Fe$_{1.78}$Se$_2$. Single crystals of RFS were grown as described elsewhere~\cite{s1gooch}. Two types of RFS were studied: one superconducting with T$_C$ onset of 31K and chemical composition Rb$_{0.93}$Fe$_{1.70}$Se$_2$; and the other insulating with composition Rb$_{0.90}$Fe$_{1.78}$Se$_2$.

As the crystals are sensitive to air, all sample preparations were preformed inside a nitrogen-flushed glove box. ARPES measurements were carried out at beamline 10.0.1 of the Advanced Light Source and beamline 5-4 of Stanford Synchrotron Radiation Lightsource using SCIENTA R4000 electron analyzers. The total energy resolution used was 25 meV or better and the angular resolution was 0.3$^\circ$C.  Single crystals were cleaved in situ at low temperatures and measured in an ultra high vacuum chamber with a base pressure better than $3x10^{-11}$ Torr.

\subsection{Discussion of orbital assignment}
The common band structure for FeSC in the Brillouin zone (BZ) corresponding to the two-Fe unit cell consists of three hole bands near the $\Gamma$ point and two electron bands at the X point~\cite{s2graser}. Under C$_4$ rotational symmetry, the \dxz~and \dyz~bands are degenerate at the in-plane high symmetry points of the BZ-$\Gamma$ and X points. At the $\Gamma$ point, this means that the \dxz~and \dyz~hole bands must be degenerate. Hence we have assigned the two lower degenerate hole bands in Fig.~\ref{fig:fig1}e to \dxzyz~and the higher one to \dxy. At the X point, C$_4$ symmetry dictates the bottom of the \dxzyz~electron band and the top of the \dxzyz~hole band be degenerate. Since the shallow electron band is separated from the hole band by more than 70meV, it is unlikely to be the \dxzyz~electron band, but is instead the \dxy~band.

There is in principle the possibility of an alternative understanding. As described in Ref.~\cite{s4Lin,s4Brouet}, the alternating arsenic atoms about the iron plane may induce parity-switching of certain orbitals when folding from the 1-Fe BZ to the 2-Fe BZ, which would swap the orbital characters of the \dxz~electron band and the \dxy~electron band at X along $\Gamma$-X direction, making the originally \dxz~electron band to be more observable under an odd polarization (Fig.~\ref{fig:fig1}c). Under this understanding, the shallow band is still the \dxz~band rather than the \dxy~band. However, for this understanding to hold, the aforementioned degeneracy of the \dxzyz~bands must be lifted to account for the 70meV gap between the shallow electron band and the hole band at X.

There are two mechanisms that could lift this degeneracy. One is spin-orbit coupling. However, in this compound, the magnitude of the spin-orbit coupling is not strong enough to cause such a large gap. Moreover, the effect of the spin-orbit coupling is stronger at the $\Gamma$ than at the X point~\cite{s3mazin}, and should cause a larger gap between the \dxzyz~hole bands at $\Gamma$, which is not observed. The second mechanism is in-plane symmetry breaking, which causes anisotropic shift of the \dxzyz~bands and hence lift the degeneracy of these two orbitals, as has been observed in the orthorhombic state of \undoped~\cite{s4Yi11}. Up to date, there has not yet been any report of in-plane symmetry breaking in the AFS superconductors. In addition, one should note that the orthorhombicity causes the existence of natural twinning in the crystals. Hence in an unstressed crystal, the ARPES signal is a combination of dispersions from both types of domains, and would see bands from $\Gamma$-X and $\Gamma$-Y directions simultaneously, as was observed in \undoped~\cite{s4Yi11}. For the microscopically phase separated AFS superconductors, such twinning effect should be even more noticeable than \undoped, but is not observed in the dispersions of AFS superconductors.

Another factor that should be considered is the comparison between $\Gamma$ and X. The flat \dxy~hole band observed at $\Gamma$ is renormalized by a factor of $\sim$10 compared to bare LDA while the \dxzyz~hole bands are renormalized by a factor of $\sim$3. When such kind of orbital-dependent renormalization is considered for the bands at X, it would bring the \dxy~electron band to be shallower than the \dxzyz~electron bands, which is more consistent with the assignment of the shallower electron band to \dxy. Further support for this assignment comes from the general consideration that a stronger mass renormalization would imply a stronger temperature dependence of the corresponding quasiparticle spectral weight, as has been shown in the main text.

To summarize, while the possibility in principle exists for other unknown mechanisms and alternative explanations, given the totality of the observations and considerations given above, the understanding of the shallow electron band as \dxy~seems to be the most consistent overall.

\subsection{Temperature cycle test}
\begin{figure}
\includegraphics[width=0.45\textwidth]{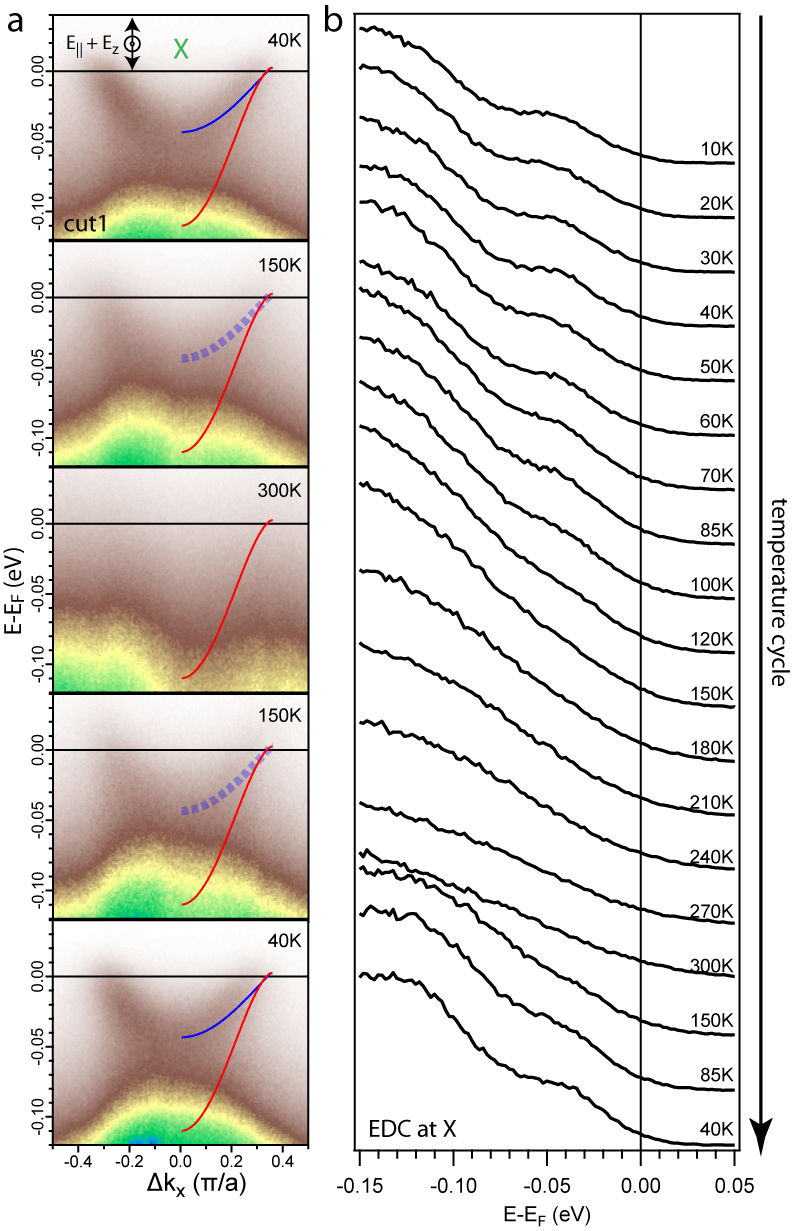}
\caption{\label{fig:figsi1}Temperature cycle test on \KFS~(a) Spectral images across X point measured at selected temperatures in the temperature cycle from 10K to 300K to 40K. Polarization geometry and photon energy is the same as that in Fig.~\ref{fig:fig2}(a). (b) EDC at X point taken in the temperature cycle, where the peak around -0.05eV is the \dxy band bottom, whose spectral weight diminishes up to 300K and recovers when cooled back down.}
\end{figure}

To test that the disappearance of the \dxy~orbital spectral weight with raised temperature is not due to sample aging, we have performed a temperature cycle test in which we cleaved the sample at 10K and took measurements as temperature is raised up to 300K and cooled back down again. Spectra images for selected temperatures as well as the EDC at X point in this temperature cycle are shown in Fig.~\ref{fig:figsi1}, where we see that the spectral weight for the \dxy~orbital dominated shallow electron band diminishes at higher temperatures, and recovers when cooled back down. This shows that the observed temperature-induced crossover is not due to any surface effects and is a robust bulk phenomenon.

\subsection{Results on \RFS}
\begin{figure*}
\includegraphics[width=0.9\textwidth]{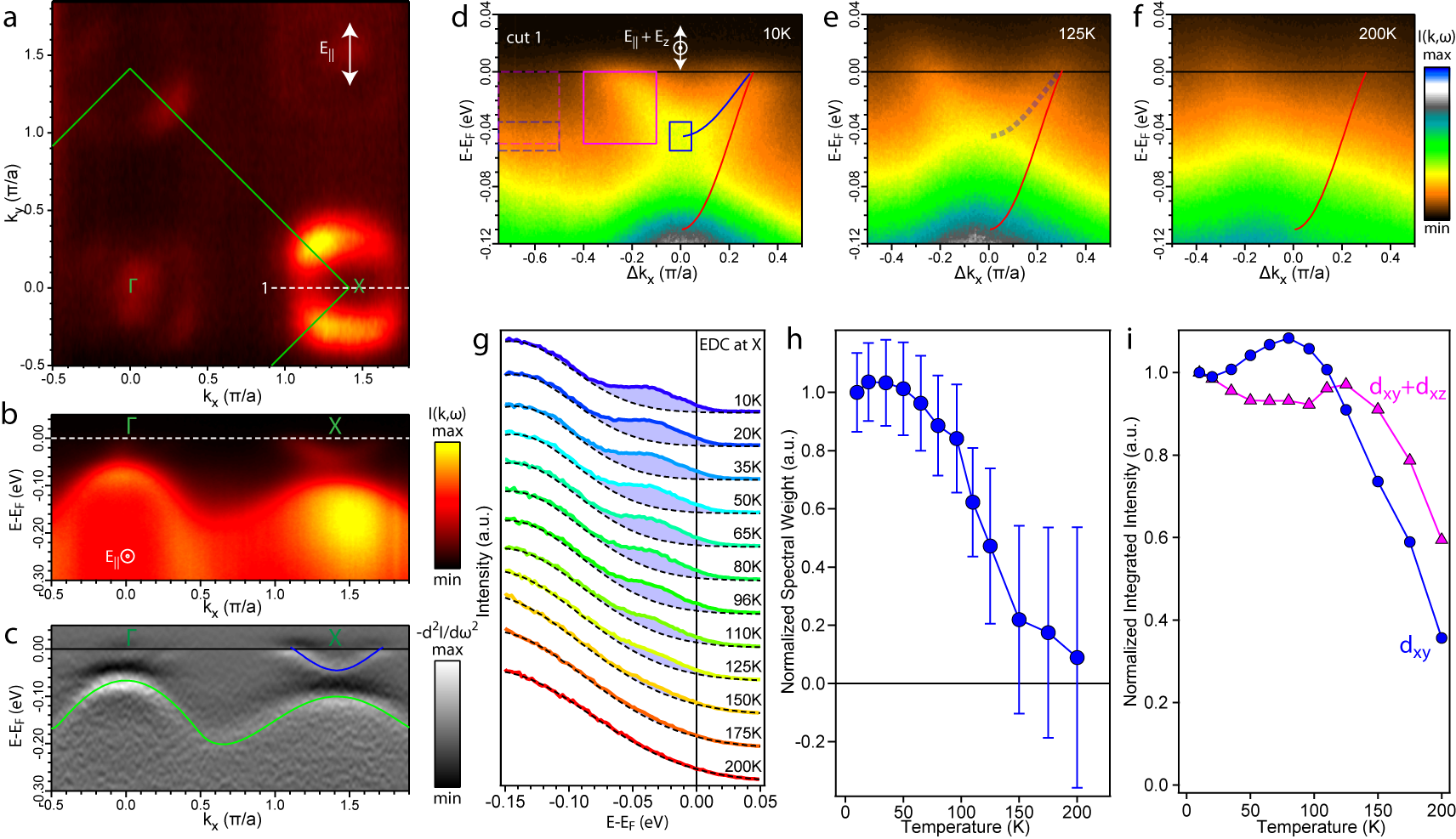}
\caption{\label{fig:figsi2}(a) Fermi surface measured at 10K, with integration window of 20meV about Fermi level. (b) Spectral image and (c) second derivative measured along $\Gamma$-X. Blue (green) curves trace the observed \dxy~(\dyz) dispersions observed under this polarization. (a)-(c) were measured with 47.5eV photons. (d)-(f) Spectral images measured across X in the same geometry as that in Fig.~\ref{fig:fig2}(a) for selected temperatures. (g) EDC at X measured in the same geometry as that for (d)-(f) for selected temperatures. Each curve is fitted to a Gaussian (dotted lines) for the background feature around -0.15eV and a Lorentzian (shaded region) for the feature around -0.05eV. (h) The integrated Lorentzian spectral weight is plotted as a function of temperature. (i) Temperature-dependence of the intensity in the regions marked by colored boxes in (d). The blue region has dominant spectral weight from the \dxy~band, whereas the magenta region is of mixed \dxy~and \dxz.}
\end{figure*}

We have done similar measurements on superconducting RFS as discussed in the main text for superconducting KFS. Fig.~\ref{fig:figsi2} summarizes the main results. RFS has similar electronic structure as KFS, including Fermi surface, band dispersions and their orbital characters, specifically there is a shallow electron band at X that is \dxy~and a deep electron band that is \dxzyz. Fig. S2d-f shows the temperature dependence of the electron bands, equivalent to Fig.~\ref{fig:fig2}(a) for the KFS. Here again we see that the shallow \dxy~band is present at low temperatures, and its spectral weight diminishes with raised temperature. Similar quantitatively analysis is also done on the RFS. Fig.~\ref{fig:figsi2}(g) shows the EDC at X point taken in the geometry of Fig.~\ref{fig:figsi2}(d)-(f), taken from 10K to 200K. Similar to the case of KFS, the $\sim$-0.05eV hump shaded by blue is the \dxy~band bottom, whose spectral weight diminishes with raised temperature. The EDCs are again fitted with a Gaussian background and a Lorentzian peak, whose integrated intensity as a function of temperature is plotted in Fig. S2h. The method of integrating intensity in boxed regions is also done on RFS. Here two regions were chosen (Fig.~\ref{fig:figsi2}d): blue box for a region dominated by \dxy~band, and magenta box for a region of mixed \dxy~and \dxz. The background for each region is taken as a box of same energy window away from dispersions, marked by dotted boxes in Fig.~\ref{fig:figsi2}(d). The integrated intensity of these two regions as a function of temperature is plotted in Fig.~\ref{fig:figsi2}(i), showing the diminishing trend of \dxy region and the slower trend of the mixed region. These behaviors in RFS are very similar to those in KFS in the main text.

\subsection{Orbital-selective Mott phase of the five-orbital Hubbard model for \KFS}
The five-orbital Hubbard model is given by $H=H_{\mathrm{kin}} + H_{\mathrm{int}}$, where $H_{\mathrm{kin}}$ and $H_{\mathrm{int}}$ respectively denote the kinetic and the on-site interaction parts of the Hamiltonian. $H_{\mathrm{int}}$ contains the intra- and inter-orbital Coulomb repulsion, as well as the Hund's rule coupling and the pair hoppings.\cite{s8yu} The corresponding coupling strengths are respectively $U$, $U^\prime$, and $J$, which satisfy $U^\prime=U-2J$.~\cite{s5castellani} For simplicity, we consider only the density-density interactions and neglect the spin-flip and pair-hopping terms. The results including these terms are qualitatively the same. The kinetic part $H_{\mathrm{kin}}$ is a tight-binding Hamiltonian, and is conveniently specified in the momentum space. In FeSCs, each unit cell contains two Fe ions. Hence, ideally the tight-binding Hamiltonian must be defined in the BZ corresponding to this two-Fe unit cell. However, the lattice symmetry of the FeSC system allows us to work in an unfolded BZ corresponding to one-Fe unit cell. Following Ref.~\cite{s6wen} and notice that the ions are invariant under the transformation $P_zT_x$ and $P_zT_y$, where $T_{x(y)}$ is the translation along $x(y)$ direction by one Fe-lattice spacing, and $P_z$ refers to the reflection about the Fe plane, we may define a pseudocrystal momentum $\bf{\tilde{k}}$ in the extended one-Fe BZ. This pseudocrystal momentum and the conventional momentum $\bf{k}$ are related by $\bf{\tilde{k}}=\bf{k}+\bf{Q}$ in \dxz~and \dyz~orbitals (where $\bf{Q}=(\pi,\pi)$), but $\bf{\tilde{k}}=\bf{k}$ in other orbitals. In the extended BZ, the tight-binding Hamiltonian reads

\begin{equation}\label{eqn:eqn2}
H_{\mathrm{kin}}=\sum\limits_{\bf{\tilde k}\alpha\beta\sigma}[\xi_{\alpha\beta}(\bf{\tilde k})+(\Delta_\alpha-\mu)\delta_{\alpha\beta}]d^{\dagger}_{\bf{\tilde k}\alpha\sigma}d_{\bf{\tilde k}\beta\sigma}.
\end{equation}

Here $\xi_{\alpha\beta}(\bf{\tilde{k}})$ is the hopping matrix in the momentum space associated with orbitals $\alpha$ and $\beta$, $\Delta_{\alpha}$ is the on-site energy reflecting the crystal field splitting, $\mu$ is the chemical potential, and $\delta_{\alpha\beta}$ is the Kronecker's delta function. The expression of $\xi_{\alpha\beta}(\bf{\tilde{k}})$ is given in Ref.~\cite{s7yu}, and it has the same form as appeared in the appendix of Ref.~\cite{s2graser}. We adopt the tight-binding parameters of Ref.~\cite{s7yu}, where they are obtained by fitting the LDA band structure of KFS. To better fit the LDA results, we have further tuned several hopping parameters by hand from their values in Ref.~\cite{s7yu}. Using the same notation as in Ref.~\cite{s2graser}, the tight-binding parameters used in this paper for KFS are listed in Table~\ref{Table:tab1}. The chemical potential corresponding to the electron filling $n$=6.15 is $\mu$=-0.365 eV.

The five-orbital Hubbard model is studied using the recently developed U(1) slave-spin mean-field method~\cite{s8yu}. In this method, a slave quantum S=1/2 spin is introduced to carry the charge degree of freedom, and the spin of the electron is carried by a fermionic spinon. This approach determines the quasiparticle spectral weight $Z_\alpha$ in each orbital. An orbital $\alpha$ is delocalized when $Z_\alpha>0$, but becomes Mott localized if $Z_\alpha=0$.

\begin{table*}
  \centering
\begin{tabular}{cccccccc}
  \hline
  \hline
    & $\alpha=1$ & $\alpha=2$ & $\alpha=3$ & $\alpha=4$ & $\alpha=5$ &   &   \\ \hline
  $\Delta_\alpha$ & -0.36559 & -0.36559 & -0.56466 & -0.05096 & -0.91583 &  & \\ \hline\hline
  $t^{\alpha\alpha}_\mu$ & $\mu=x$ & $\mu=y$ & $\mu=xy$ & $\mu=xx$ & $\mu=xxy$ & $\mu=xyy$ & $\mu=xxyy$ \\ \hline
  $\alpha=1$ & -0.11475 & -0.38868 & 0.20881 & -0.04557 & -0.00866 & -0.03143 & 0.01899\\ \hline
  $\alpha=3$ & 0.32523 &  & -0.09783 & -0.00537 &  &  &  \\ \hline
  $\alpha=4$ & 0.20633 &  & 0.09682 & -0.07525 & -0.02189 &  & 0.00423 \\ \hline
  $\alpha=5$ & -0.0427 &  &  & 0.01117 & 0.00177 &  & -0.01349 \\ \hline\hline
  $t^{\alpha\beta}_\mu$ & $\mu=x$ & $\mu=xy$ & $\mu=xxy$ & $\mu=xxyy$ &  &  &  \\ \hline
  $\alpha\beta=12$ &  & 0.10161 & -0.02017 & 0.03273 &  &  &  \\ \hline
  $\alpha\beta=13$ & -0.31447 & 0.06225 & 0.0103 &  &  &  &  \\ \hline
  $\alpha\beta=14$ & 0.13785 & -0.03105 & 0.0104 &  &  &  &  \\ \hline
  $\alpha\beta=15$ & -0.04825 & -0.10096 &  & -0.01204 &  &  &  \\ \hline
  $\alpha\beta=34$ &  &  & -0.04795 &  &  &  &  \\ \hline
  $\alpha\beta=35$ & -0.30966 &  & -0.01498 &  &  &  &  \\ \hline
  $\alpha\beta=45$ &  & -0.08359 &  & -0.00766 &  &  &  \\ \hline
  \hline
\end{tabular}
\caption{\label{Table:tab1}Tight-binding parameters of the five-orbital model for \KFS. Here we use the same notation as in Ref~\cite{s2graser}. The orbital index $\alpha$=1,2,3,4,5 correspond to \dxz, \dyz, \dxxyy, \dxy, and \dz~orbitals, respectively. The units of the parameters are eV. }
\end{table*}

\begin{figure*}
\includegraphics[width=0.9\textwidth]{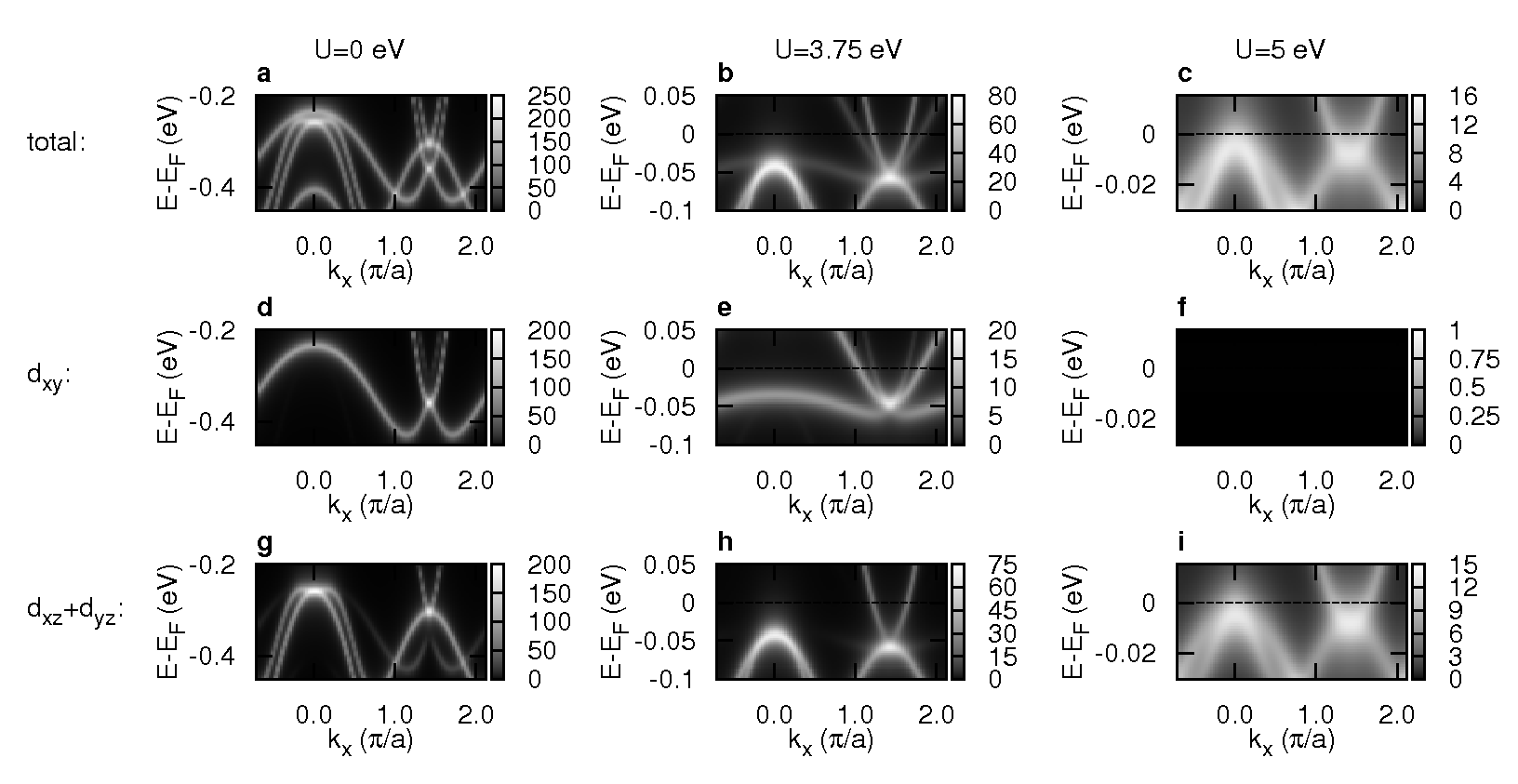}
\caption{\label{fig:figsi3}Calculated quasiparticle spectral weight functions for various U. (a) Orbitally resolved quasiparticle spectral weight functions along the $\Gamma$ to X direction of the Brillouin zone at J/U=0.15, n=6.15, T=10 K, and three different U values (in three columns). The total spectral weights contributed from all five orbitals are shown in the first row; the spectral weights from the \dxy~and \dxz+\dyz~orbitals are respectively shown in the second and the third row.}
\end{figure*}

To compare with the ARPES data, the coherent part of the orbital resolved spectral function $A_\alpha(\bf{k},E)$ in the folded BZ corresponding to the two-Fe unit cell is further calculated by convoluting the slave-spin and spinon Green's functions and writing them in the Lehmann representation via the following formula

\begin{equation}\label{eqn:eqn5}
A_\alpha(\bf{k},E) = \sum\limits_{\alpha\nu} Z_\alpha| U^{\alpha\nu}_{\bf{k}}|^2 \delta(E-\varepsilon_{\nu\bf{k}}).
\end{equation}

Here, $\varepsilon_{\nu\bf{k}}$ and $U^{\alpha\nu}_{\bf{k}}$ are respectively the $\nu$'s eigenenergy and eigenvector of the hopping matrix $\xi_{\alpha\beta}(\bf{k})$. This allows us to determine the orbital character of each band near the Fermi level. In Fig.~\ref{fig:figsi3}, we show the spectral functions near the Fermi level in the five-orbital model at n=6.15 and T=10 K for three different U values. Compared to the U=0 band structure, the bands are strongly renormalized by the interactions. Moreover, with increasing U, the spectral weight of the bands with a \dxy~orbital character is reduced (Fig.~\ref{fig:figsi3}(b)) and eventually goes away (Fig.~\ref{fig:figsi3}(c)). The bands with the \dxzyz~character, on the other hand, still have nonzero spectral weights. This behavior clearly indicates an interaction driven transition to an OSMP.

Several factors favor stabilizing the OSMP.\cite{s15yu} Firstly, from the orbitally-projected density of states of the non-interacting bands, the width of the \dxy~orbital is narrower than that of the other Fe 3d orbitals, especially the \dxzyz~orbitals. This factor recalls the mechanism for the OSMP initially proposed for the Ca$_{2-x}$Sr$_x$RuO$_4$ system~\cite{s10anisimov}. Secondly, when the Hund's coupling is sufficiently strong compared to the (nonzero) splitting between the \dxy~and \dxzyz~orbitals, the high-spin configuration is favored, and, due to the crystal level splitting, the \dxy~orbital is non-degenerate and located at a higher energy than other orbitals. As a result, for a range of densities, the \dxy~orbital is kept at half-filling while the \dxzyz~orbitals are more than (though still close to) half filled. The non-degenerate \dxy~orbital has a lower repulsion threshold for the Mott transition than the doubly degenerate \dxzyz~orbitals within a wide range of Hund's coupling. This picture has some connection with the one studied in a different regime for a three-orbital model away from half-filling~\cite{s11demedici}. As a combined effect of these factors, the \dxy~orbital is more strongly localized than the \dxzyz~orbitals.

\subsection{Comparison of insulating, intermediate and superconducting compounds}

\begin{figure*}
\includegraphics[width=0.9\textwidth]{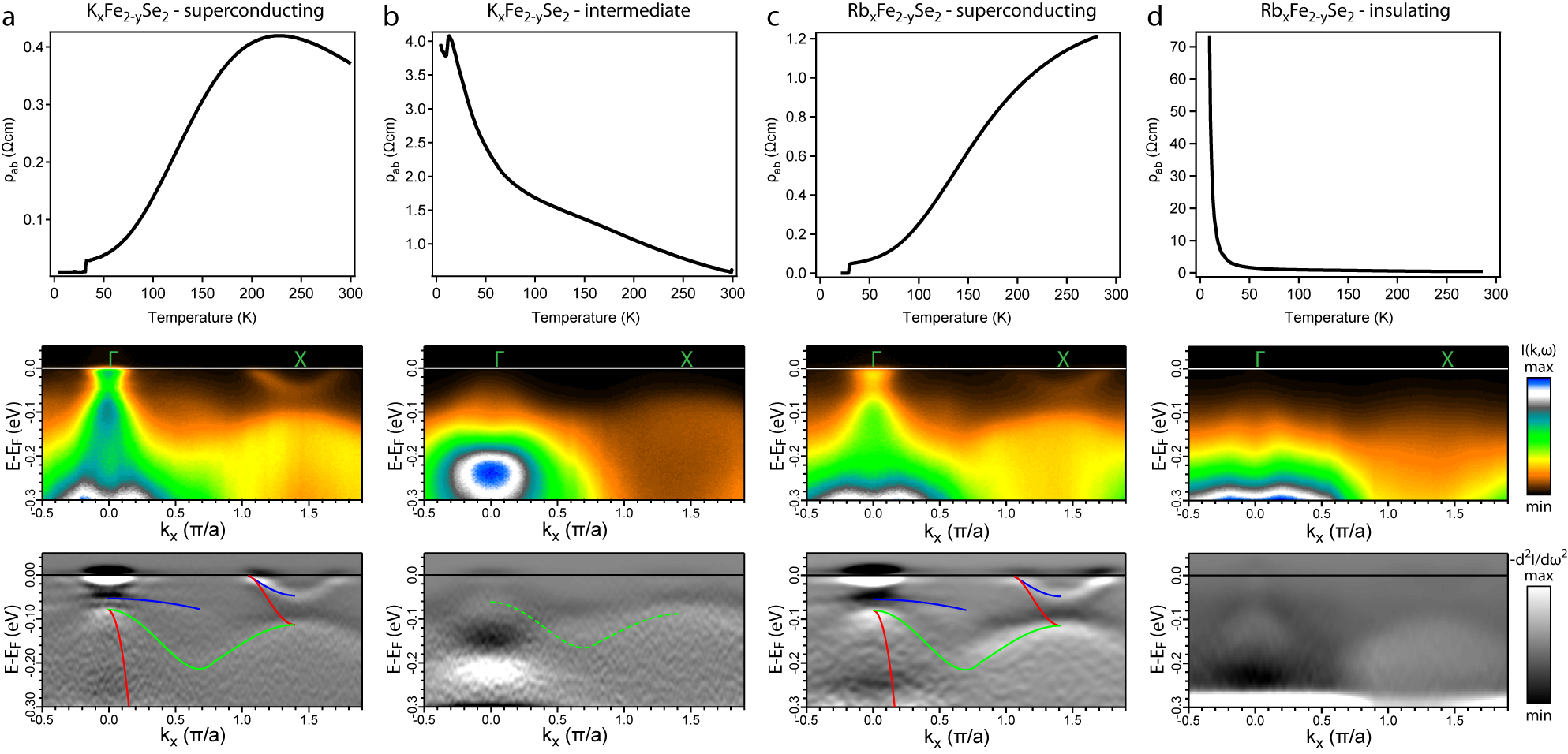}
\caption{\label{fig:figsi4}Comparison of superconducting, intermediate and insulating compounds (a) Resistivity (top), ARPES spectral images on the $\Gamma$-X high symmetry cut (middle) and its second derivative plot (bottom) for superconducting K$_{0.76}$Fe$_{1.72}$Se$_2$, taken at 10K. Same measurements are shown for (b) intermediate $K_{0.76}$Fe$_{1.78}$Se$_2$ at 10K, (c) superconducting Rb$_{0.93}$Fe$_{1.70}$Se$_2$ at 10K, and (d) insulating K$_{0.90}$Fe$_{1.78}$Se$_2$ at 50K. All ARPES data were taken with 26eV photons in the same polarization geometry as that for Fig.~\ref{fig:fig1}(d).}
\end{figure*}

\begin{figure}
\includegraphics[width=0.5\textwidth]{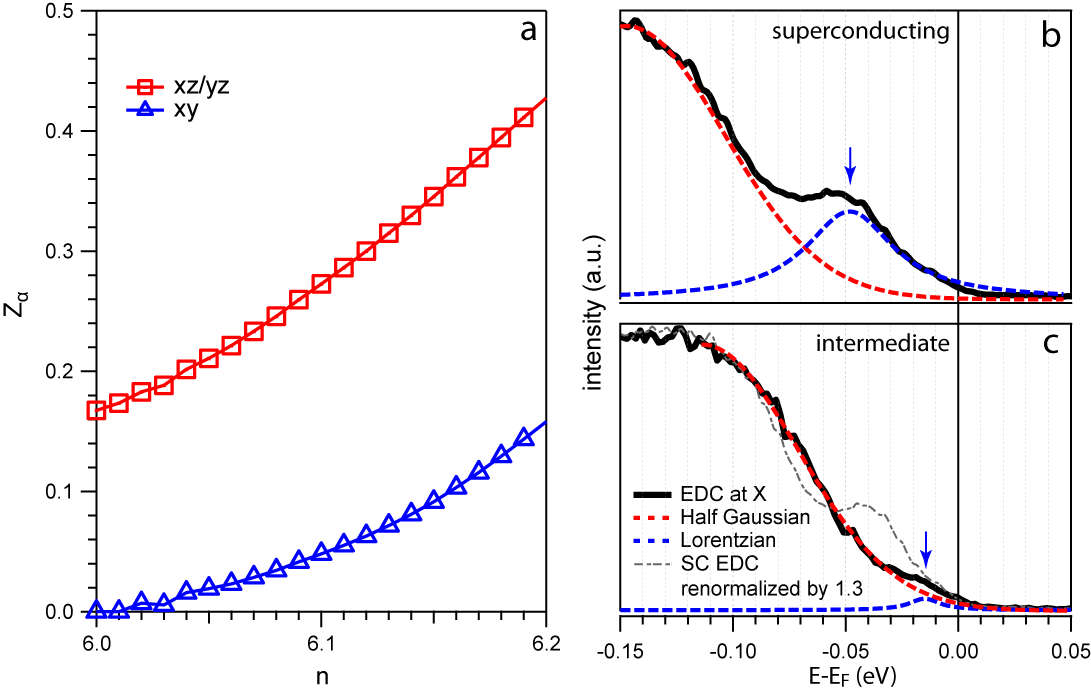}
\caption{\label{fig:figsi5}Comparison of superconducting and intermediate KFS (a) Calculated orbital-resolved quasiparticle spectral weight for U=3.75eV and T=30K as a function of electron filling, n, for \dxzyz~(red squares) and \dxy~(blue triangles). (b) Measured EDC at X (black solid curve) from superconducting KFS from Fig.~\ref{fig:figsi4}(a), fitted to a Gaussian plus a Lorentzian, plotted in red and blue, respectively. (c) Same as (b) but for the intermediate KFS sample from Fig. S4b, with the EDC from the superconducting sample renormalized by 1.3 in energy for comparison (gray dotted curve). Blue arrows point to the positions of the \dxy~band bottom.}
\end{figure}

Here, we compare the measured electronic structure of three kinds of phases in KFS and RFS. The four compounds shown in Fig.~\ref{fig:figsi4} have compositions determined by energy dispersive X-ray spectroscopy to be (a) K$_{0.76}$Fe$_{1.72}$Se$_2$, (b) $K_{0.76}$Fe$_{1.78}$Se$_2$, (c) Rb$_{0.93}$Fe$_{1.70}$Se$_2$, and (d) K$_{0.90}$Fe$_{1.78}$Se$_2$. From resistivity data shown, (a) and (c) are superconducting, (d) is insulating, and (b) shows a behavior somewhat between superconducting and insulating, which we shall call intermediate. Its resistivity is most like the compound suggested to be semiconducting in a previous report~\cite{s12chen}, which suggests that both insulating and metallic regions exist in these samples. All compounds were measured in the same experimental geometry. The two superconducting samples exhibit similar dispersions as discussed in previous section. The insulating sample shows negligible spectral weight towards \EF~and no well defined dispersions near \EF, as expected of an insulator. The interesting case is the intermediate sample, which has resolvable dispersions, but different from that of the superconducting samples. Firstly, we see that the most well-resolved band is the \dyz~band, which is traced in green in the second derivative plot in the bottom panel. Its bandwidth is approximately renormalized by a factor of 1.3 compared to its equivalent in the superconducting samples. Secondly, from the EDC taken at the X point from both the superconducting KFS and the intermediate KFS (Fig.~\ref{fig:figsi5}(b)-(c)), we see that the peak around -0.05eV indicating the \dxy~electron band bottom in the superconducting sample (Fig.~\ref{fig:figsi5}(b)) becomes a smaller but discernible shoulder closer to \EF~in the intermediate sample (Fig.~\ref{fig:figsi5}(c)). Even after a renormalization of 1.3 to account for the increased renormalization of the \dyz~band (gray dotted line in Fig.~\ref{fig:figsi5}(c)) as discussed above, we see that the \dxy~band in the intermediate compound still needs a further renormalization going from the superconducting to the intermediate sample (Fig.~\ref{fig:figsi5}(c)). Also, the relative spectral weight of the \dxy~orbital to that of \dyz~is much more reduced in the intermediate sample. Both the further renormalization of the bandwidth and the reduction of spectral weight of the \dxy~band indicate that the intermediate sample is even closer to the OSMP than the superconducting compounds at low temperatures.

Assuming the same interaction strengths, the various phases in the KFS and RFS compounds can be understood in terms of an interplay between the electron doping and vacancy order. In the vacancy disordered case, our calculation (Fig.~\ref{fig:figsi5}(a)) identifies a doping-induced transition to an OSMP near n=6.02 per Fe. Hence the undoped system with n=6 is already in an OSMP. The vacancy order further drives it through a Mott transition in all orbitals to a Mott insulator~\cite{s13yu,s14craco}. This accounts for the absence of spectral weights near \EF~in the insulating RFS. For the superconducting samples and the intermediate KFS compound, the phases giving ARPES signals are likely to be vacancy disordered. We therefore interpret them as corresponding to two vacancy disordered phases at two different electron densities (n=6.15-6.25 for the superconducting samples, and n=6.05-6.10 for the metallic phase in the intermediate KFS). In addition, we emphasize that the high temperature state of the superconducting region is intrinsically different from the vacancy-ordered insulating phase. Rather, the superconducting, intermediate, and insulating phases likely have increasing correlation as they may be located close to an OSMP, just at the boundary of an OSMP, and in a Mott insulating phase, respectively.


\begin{thebibliography}{99}

\bibitem{1LeePA06} P.A. Lee, N. Nagaosa, and X.-G. Wen, Rev. Mod. Phys. 78, 17-85 (2006).
\bibitem{2LuDH08} D.H. Lu $et~al.$, Nature 455, 81 (2008).
\bibitem{3YangWL09} W.L. Yang W. L. $et~al.$, Phys. Rev. B 80, 014508 (2009).
\bibitem{4Qazibash09} M.M. Qazilbash $et~al.$, Electronic correlations in the iron pnictides. Nat. Phys. 5, 647 (2009).
\bibitem{5YinZP11} Z.P. Yin, K. Haule, and G. Kotliar, Nat. Mat. 10, 932-935 (2011).
\bibitem{6Liu10} T.J. Liu $et~al.$, Nat. Mat. 9, 718-720 (2010).
\bibitem{7Guo10} J. Guo $et~al.$, Phys. Rev. B 82, 180520(R) (2010).
\bibitem{8Krzton11} A. Krzton-Maziopa $et~al.$, J. Phys.: Condens. Matter 23, 052203 (2011).
\bibitem{9Li11} C.-H. Li, B. Shen, F. Han, X. Zhu, and H.-H. Wen, Phys. Rev. B 83, 184521 (2011).
\bibitem{10Fang11} M.-H. Fang $et~al.$, Europhys. Lett. 94, 27009 (2011).
\bibitem{11Wang11} H.-D. Wang $et~al.$, Europhys. Lett. 93, 47004 (2011).
\bibitem{12Wei11} W. Bao $et~al.$, Chin. Phys. Lett. 28, 086104 (2011).
\bibitem{13Yi11} M. Yi, $et~al.$, PNAS 108, 6878 (2011).
\bibitem{14chubukov12} A.V. Chubukov, Mat. Phys. 3, 52 (2012).
\bibitem{15si08} Q. Si, and E. Abrahams, Phys. Rev. Lett. 101, 076401 (2008).
\bibitem{16Yu11} R. Yu, and Q. Si, Phys. Rev. B 84, 235115 (2011).
\bibitem{17Yu11b} R. Yu, J.-X. Zhu, and Q. Si, Phys. Rev. Lett. 106, 186401 (2011).
\bibitem{18Zhou11} Y. Zhou, D.-H. Xu, F.-C. Zhang, and W.-Q. Chen, Europhys. Lett. 95, 17003 (2011).
\bibitem{19craco11} L. Craco, M.S. Laad, and S. Leoni, Phys. Rev. B 84, 224520 (2011).
\bibitem{20anisimov02} V. Anisimov $et~al.$, Eur. Phys. J. B 25, 191 (2002).
\bibitem{21demedici} L. de' Medici, S.R. Hassan, M. Capone, and X. Dai, Phys. Rev. Lett. 102, 126401 (2009).
\bibitem{22You11} Y.-Z. You, F. Yang, S.-P. Kou, and Z.-Y. Weng, Phys. Rev. B 84, 054527 (2011).
\bibitem{23moon10} S. J. Moon $et~al.$, Phys. Rev. B 81, 205114 (2010).
\bibitem{24zhang11} Y. Zhang $et~al.$, Nat. Mat. 10, 273-277 (2011).
\bibitem{25qian11} T. Qian $et~al.$, Phys. Rev. Lett. 106, 187001 (2011).
\bibitem{26mou11} D. Mou $et~al.$, Phys. Rev. Lett. 106, 107001 (2011).
\bibitem{27nekrasov11} I.A. Nekrasov, and M.V. Sadovskii, JETP Lett. 93, 166-169 (2011).
\bibitem{28graser10} S. Graser $et~al.$, Phys. Rev. B 81, 214503 (2010).
\bibitem{30yu} R. Yu, and Q. Si, Phys. Rev. B 86, 085104 (2012).
\bibitem{31demedici05} L. de'Medici, L., A. Georges, and S. Biermann, Phys. Rev. B 72, 205124 (2005).
\bibitem{29yu} R. Yu, and Q. Si, unpublished (2012).
\bibitem{32li12} W. Li $et~al.$, Nat. Phys. 8, 126-130 (2012).
\bibitem{33chen11} F. Chen $et~al.$, Phys. Rev. X 1, 021020 (2011).

\end{thebibliography}

\begin{thebibliography}{99}
\bibitem{s1gooch} M. Gooch $et~al.$, Phys. Rev. B 84, 184517 (2011).
\bibitem{s2graser} S. Graser, T.A. Maier, P.J. Hirschfeld, and D.J. Scalapino, New J. Phys. 11, 025016 (2009).
\bibitem{s3mazin} I.I. Mazin. private communications.
\bibitem{s4Yi11} M. Yi $et~al.$, PNAS 108, 6878 (2011).
\bibitem{s4Lin} C.-H. Lin $et~al.$, Phys. Rev. Lett. 107, 257001 (2011).
\bibitem{s4Brouet} V. Brouet $et~al.$, arXiv:1205.4513.
\bibitem{s5castellani} C. Castellani, C.R. Natoli, and J. Ranninger, Phys. Rev. B 18, 4945 (1978).
\bibitem{s6wen} X.-G. Wen, and P.-A. Lee, Phys. Rev. B 78, 144517 (2008).
\bibitem{s7yu} R. Yu $et~al.$, arXiv:1103.3259.
\bibitem{s8yu} R. Yu, and Q. Si, Phys. Rev. B 86, 085104 (2012).
\bibitem{s9yu} R. Yu, and Q. Si, Phys. Rev. B 84, 235115 (2011).
\bibitem{s15yu} R. Yu, and Q. Si, unpublished (2012).
\bibitem{s10anisimov} V. Anisimov $et~al.$, Eur. Phys. J. B 25, 191 (2002).
\bibitem{s11demedici} L. de' Medici, S.R. Hassan, M. Capone, and X. Dai, Phys. Rev. Lett. 102, 126401 (2009).
\bibitem{s12chen} F. Chen $et~al.$, Phys. Rev. X 1, 021020 (2011).
\bibitem{s13yu} R. Yu, J.-X. Zhu, and Q. Si, Phys. Rev. Lett. 106, 186401 (2011).
\bibitem{s14craco} L. Craco, M.S. Laad, and S. Leoni, Phys. Rev. B 84, 224520 (2011).

\end{thebibliography}
\end{document}